# A Global Perspective with Updated Constraints on the Ultra-hot Jupiter WASP-19b: Atmospheric Properties and Stellar Activity

Abigail A. Tumborang[1,2], Jessica J. Spake[3,12], Heather A. Knutson[3], Megan Weiner Mansfield[4,12], Kimberly Paragas[3],
Billy Edwards[5,6], Tiffany Kataria[7], Thomas M. Evans-Soma[8,9], Nikole K. Lewis[10], and Gilda E. Ballester[11]
[1] School of Physics and Astronomy, University of St. Andrews, St. Andrews, Scotland, KY16 9SS, UK
[2] Kapteyn Astronomical Institute, University of Groningen, 9700 AV Groningen, The Netherlands
[3] Division of Geological and Planetary Sciences, California Institute of Technology, Pasadena, CA 91125, USA
[4] Department of Astronomy and Steward Observatory, University of Arizona, 933 North Cherry Avenue, Tucson, AZ 85721-0065, USA
[5] SRON, Netherlands Institute for Space Research, Niels Bohrweg 4, NL-2333 CA, Leiden, The Netherlands
[6] Department of Physics and Astronomy, University College London, Gower Street, London, WC1E 6BT, UK
[7] Jet Propulsion Laboratory, 4800 Oak Grove Drive, Pasadena, CA 91109, USA
[8] School of Information and Physical Sciences, University of Newcastle, Callaghan, NSW 2308, Australia
[9] Max Planck Institute for Astronomy, Königstuhl 17, D-69117 Heidelberg, Germany
[10] Department of Astronomy and Carl Sagan Institute, Cornell University, 122 Sciences Drive, Ithaca, NY 14853, USA
[11] Lunar & Planetary Laboratory, Department of Planetary Sciences, University of Arizona, Tucson, AZ 85721, USA
*Received 2024 April 10; revised 2024 October 2; accepted 2024 October 11; published 2024 December 2*

## Abstract

We present a detailed reanalysis of the atmospheric properties of WASP-19b, an ultra-hot Jupiter (1.14 $M_{\rm Jup}$, 1.41 $R_{\rm Jup}$) orbiting an active Sun-like star every 0.79 day. We reanalyze a transit and secondary eclipse of WASP-19b observed by the Hubble Space Telescope's Wide Field Camera 3 spectrograph (1.1–1.7 $\mu$m). When combined with Spitzer photometry at longer wavelengths, our analyses indicate the presence of water absorption features in both the planet's transmission and emission spectra, consistent with results from previously published studies. We jointly fit WASP-19b's dayside emission and transmission spectra with a retrieval model in order to constrain its atmospheric composition, and explore the effect of stellar activity on its transmission spectrum in greater depth. We also compare our dayside emission spectrum to predictions from a general circulation model, and conclude that magnetic drag appears to be relatively unimportant in shaping WASP-19b's atmospheric circulation. Lastly, we compare the size of WASP-19b's dayside water absorption feature to the population of hot Jupiters with similar measurements, and show that it is located in the transitional irradiation regime where temperature inversions first begin to emerge. As in previous studies, we find that the current observations provide relatively weak constraints on this planet's atmospheric properties. These constraints could be significantly improved by the addition of spectroscopically resolved observations at longer wavelengths with JWST/NIRSpec PRISM.

*Unified Astronomy Thesaurus concepts:* Exoplanet atmospheric composition (2021); Exoplanet atmospheres (487); Transmission spectroscopy (2133); Light curves (918); Hubble Space Telescope (761)

## 1. Introduction

Analyses of the transmission and emission spectra from transiting exoplanets provide a wealth of information on the properties of their atmospheres. While transmission spectroscopy probes the day–night terminator of an exoplanet's atmosphere at shallower pressures (≈1–10 mbar), emission spectroscopy allows us to characterize the planet's dayside atmosphere at deeper pressures (≈100 mbar). By combining these two viewing geometries, we can probe gradients in the atmospheric composition and cloud properties driven by the day–night temperature gradients on these highly irradiated, tidally locked planets (D. Charbonneau et al. 2002; D. Deming et al. 2005; S. Redfield et al. 2008; L. Kreidberg et al. 2014a; D. K. Sing et al. 2016; L.-P. Coulombe et al. 2023; Z. Rustamkulov et al. 2023).

Ultra-hot Jupiters are the most highly irradiated subset of close-in gas giant planets, with equilibrium temperatures typically ranging between 2200 and 4000 K (T. J. Bell & N. B. Cowan 2018; V. Parmentier et al. 2018). For these planets, the dayside atmosphere can be hot enough to dissociate and ionize neutral species present at the terminator, resulting in reduced absorption from water and increased continuum opacity from H$^-$ as well as a variety of metal species. The Hubble Space Telescope (HST) Wide Field Camera 3 (WFC3) instrument has surveyed several dozen hot Jupiters in emission to date, providing us with a population-level view of how these chemical changes systematically alter the dayside emission spectra of hot Jupiters as a function of increasing equilibrium temperature (J. D. Lothringer et al. 2018; M. Mansfield et al. 2021; Q. Changeat et al. 2022). M. Mansfield et al. (2021) analyzed the dayside emission spectra of 19 hot Jupiters measured by WFC3 using the G141 grism, and performed atmospheric retrievals to quantify empirical trends in the strength of the water feature at 1.4 $\mu$m as a function of dayside temperature. They found that hot Jupiters with equilibrium temperatures cooler than around 2000 K displayed stronger water absorption features, while those with equilibrium temperatures hotter than 2000 K tended to show flatter emission spectra, or weak water emission due to thermal inversions. The presence of temperature inversions in the atmospheres of more highly irradiated hot Jupiters has been attributed to the presence of gas-phase molecules such as TiO and VO, while the dissociation of water molecules at the highest temperatures results in lower water opacities. A study by Q. Changeat et al. (2022) on the thermal structures of a population of exoplanets, including WASP-19b, echoed these results and placed WASP-19b at the upper boundary of planets that exhibit water absorption features. With an equilibrium temperature of 2100 K, WASP-19b lies at the cusp of the transition from hot to ultra-hot Jupiters, making it a particularly useful system for understanding the factors that dictate the transition from water absorption to water emission in hot Jupiter atmospheres. WASP-19b is a highly favorable target for atmospheric studies, and is one of the few ultra-hot Jupiters known to orbit an active star (L. Hebb et al. 2009; M. Lendl et al. 2013). The G8V host star in this system has a measured rotation period of $10.7 \pm 0.5$ days and a photometric variability of 1.8% (peak to peak) at optical/infrared wavelengths (M. Lendl et al. 2013; I. Wong et al. 2016, 2020).

---

[12] NHFP Sagan Fellow.

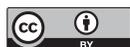

WASP-19b's atmosphere has been extensively characterized at wavelengths ranging from the optical to mid-infrared. The first study of WASP-19b by C. M. Huitson et al. (2013) measured the planet's 1.1–1.6 $\mu$m transmission spectrum using HST WFC3 and reported a detection of water absorption. When combined with optical transmission spectra measured using the Space Telescope Imaging Spectrograph (STIS) and photometric Spitzer transit observations at longer wavelengths, these data imply a water abundance of $\log(H_2O) \approx -4$ near the day–night terminator (D. K. Sing et al. 2016; J. K. Barstow et al. 2017; A. Pinhas et al. 2019). This planet's spectroscopically resolved near-infrared thermal emission has also been measured using ground-based observations (1.25–2.35 $\mu$m; J. L. Bean et al. 2013), as well as photometrically by Spitzer at longer wavelengths (3–10 $\mu$m; D. R. Anderson et al. 2013). A. S. Rajpurohit et al. (2020) interpreted these data by constructing synthetic emission spectrum models using the PHOENIX atmosphere modeling code (F. Allard et al. 2012; A. S. Rajpurohit et al. 2012) and found evidence for a weak water absorption feature at 1.4 $\mu$m. The best constraints to date on WASP-19b's atmospheric temperature profile and dayside water abundance come from HST WFC3 secondary-eclipse data presented in Q. Changeat et al. (2022), who performed both free and equilibrium chemistry retrievals on the HST and Spitzer observations. WASP-19b's day–night temperature gradient has also been measured with 3.6 and 4.5 $\mu$m Spitzer phase-curve observations (I. Wong et al. 2016), and its optical phase curve has been measured using observations from TESS (with detector bandpass spanning 600–100 nm; I. Wong et al. 2020).

In this work, we reanalyzed archival HST WFC3 observations of a primary and secondary eclipse of WASP-19b using the G141 grism and performed atmospheric retrievals on the resulting data in conjunction with other data sets. WASP-19b has been observed by three separate HST programs to date: a primary transit was observed with WFC3 in 2011 July (ID: GO-12181, PI: Deming) with independent reductions published in D. Deming et al. (2011), C. M. Huitson et al. (2013), D. K. Sing et al. (2016), and B. Edwards et al. (2023); three additional primary transits were observed with STIS in 2012 May (ID: GO-12473, PI: Sing) and published in C. M. Huitson et al. (2013) and D. K. Sing et al. (2016); and a third primary transit was observed with WFC3 along with a secondary eclipse in 2014 June (ID: GO-13431, PI: Huitson). The transit from this last program was published in a population study by B. Edwards et al. (2023), and then analyzed jointly with the secondary-eclipse data in another population study by Q. Changeat et al. (2022). The data we use (ID: GO-13431) were obtained as part of a discontinuous, partial phase-curve observation that was split into three visits spanning five HST orbits each. The data proved difficult to stitch together into a single phase curve because WASP-19 is an active star, and the flux offsets between visits were degenerate with the phase-curve amplitude. As a result, we focused on measuring an emission and transmission spectrum for WASP-19b.

We expanded on published studies of this planet by running retrievals on different subsets of the data (i.e., emission spectrum, transmission spectrum, with and without Spitzer data) to explore the effect that each of these data sets has on the retrieved atmospheric properties, and by quantifying the extent to which repeated epochs of optical and near-infrared observations differ as a result of stellar variability. We also calculated an empirical water-feature strength for the first time and used this to place WASP-19b within M. Mansfield et al.'s (2021) hot Jupiter spectral sequence. Lastly, we compared our dayside emission spectrum to predictions from general circulation models (GCMs) for this planet from T. Kataria et al. (2016) and I. Wong et al. (2016), and placed corresponding constraints on the importance of magnetic drag in WASP-19b's partially ionized atmospheric layers. The structure of this paper is as follows: Section 2 details how the observations of the primary transit and secondary eclipse were taken and the data-reduction process, Section 3 describes the light-curve fits, Section 4 presents our atmosphere models, and Section 5 places our results in the context of previously published studies of WASP-19b and other similar planets. We summarize our results in Section 6.

## 2. Observations and Data Reduction

The data presented here were taken as part of the HST program GO-13431 (PI: Huitson). This program observed WASP-19b using HST WFC3's G141 grism (1.080–1.690 $\mu$m), which has a resolving power of $R = 130$ at 1.4 $\mu$m and a dispersion of 0.005 $\mu$m pixel$^{-1}$. We observed WASP-19b on three separate visits: the first on UT 2014 June 10, the second on UT 2014 June 12, and the third on UT 2014 June 13. Each visit contained a total of 148 exposures spanning five HST orbits, with an exposure length of 46.7 s. We used HST's spatial scan mode and a scan rate of 0.22 pixels s$^{-1}$, which spread WASP-19's spectrum over 10 pixels in the spatial axis. We used the SPARS25 sampling sequence with three nondestructive reads per exposure (NSAMP = 3). The secondary eclipse occurred during visit 1 on June 10, and the primary transit during visit 2 on June 12. For this analysis, we did not use visit 3, which covered the orbital phases of 0.04–0.28 (i.e., between primary transit and secondary eclipse).

Following a similar methodology to T. Mikal-Evans et al. (2019), we extracted the spectra from raw data frames using a Python pipeline built specifically for this purpose. First, we loaded the .ima files, which already had been calibrated with flat-fielding, bias subtraction, and nonlinearity corrections. Then, we obtained the flux from our target star by measuring the difference between each nondestructive read over the duration of the exposure time, and then subtracting the background flux, obtained from calculating the median pixel count from a blank region chosen to be the background. By taking the differences between the successive nondestructive reads, we could correct for the noise caused by cosmic rays, and drifts along the dispersion axis.

In order to find the optimal aperture radius, we extracted the spectra with different aperture radii, ranging between 10 and 40 pixels in 5 pixel increments. For each aperture radius, we fitted the white-light light curves (explained in further detail in Section 3.1), and calculated the standard deviation of the residuals. We found that the aperture radius with the smallest residual standard deviation for the secondary eclipse was 25 pixels, and 20 pixels for the primary transit.

For each frame, we calculated the best-fit position of the target star, and set all pixel values outside the optimized aperture to zero. Following this process, we obtained a wavelength solution for each extracted 1D spectrum by cross-correlating them with a PHOENIX stellar model whose parameters best matched the star WASP-19 ($\log(g) = 4.5$, $T_{\text{eff}} = 5500$ K; T.-O. Husser et al. 2013).

## 3. Light-curve Fitting

### 3.1. Secondary-eclipse White-light Light Curve

Using the 1D spectra generated from visit 1, we obtained a white-light light curve for the secondary eclipse by integrating individual spectra across the full wavelength range. The resulting light curve showed hook-shaped systematics every orbit, with the first orbit being the worst affected. This phenomenon is prevalent in HST WFC3 data (Z. K. Berta et al. 2012; L. Kreidberg et al. 2014b; H. R. Wakeford et al. 2016) and is most likely caused by charge trapping in the detectors (Y. Zhou et al. 2017). We chose to exclude the first orbit from our analyses, as is common practice (A. Tsiaras et al. 2018; T. Mikal-Evans et al. 2019; J. J. Spake et al. 2021). In addition to that, we also dropped another orbit farthest from the transit event, to make it easier to fit a linear base trend onto the light curves.

We fit the white-light eclipse signal using the `batman` software package (L. Kreidberg et al. 2015), with the eclipse depth and transit mid-time $t_0$ as free parameters. Considering WASP-19b's proximity to its host star and lack of outer companions that could maintain a nonzero eccentricity in the face of ongoing tidal circularization, we do not fit for eccentricity and thus set it to zero. Allowing $a/R_{\text{star}}$ and the inclination to vary as free parameters does not affect the results of our fits, so we have elected to fix these values to the published literature values as shown in Table 1. To correct for the systematic trends within each orbit, we considered a range of systematics models including a Gaussian process model and an exponential model. For our exponential model, we used

$$M(t) = E(t)(cs + vt_{\text{vis}})(1 - e^{-r_1 t_{\text{orb}} - r_2}) \quad (1)$$

(M. Mansfield et al. 2021), where $E(t)$ is the eclipse model calculated by `batman`, $cs + vt_{\text{vis}}$ is the linear base model dependent on visit time ($t_{\text{vis}}$) with a constant term $cs$ and a linear term $v$. $r_1$ is the scaling factor and $r_2$ is the constant term for the exponential model, which is dependent on the HST orbital phase $t_{\text{orb}}$. For the Gaussian process model, we used the Python library for GP regression `george` (S. Ambikasaran et al. 2015) with a Matern $\nu = 3/2$ kernel, with the HST orbital phase $\phi$, the position in the dispersion direction $y$, and the position in the spatial direction on the detector $x$ as the input variables as described in T. M. Evans et al. (2013, 2018) and

**Table 1**
System Parameters Used in This Work

| Parameters | WASP-19 | Source |
|---|---|---|
| $M_*$ | $0.965^{+0.091}_{-0.095}\ M_\odot$ | P. Cortés-Zuleta et al. (2020) |
| $R_s$ | $1.006^{+0.031}_{-0.034}\ R_\odot$ | P. Cortés-Zuleta et al. (2020) |
| $\log(g)$ | $4.417^{+0.020}_{-0.021}$ | P. Cortés-Zuleta et al. (2020) |
| $T_{\text{eff}}$ | $5616^{+66}_{-65}$ K | P. Cortés-Zuleta et al. (2020) |
| $M_p$ | $1.154^{+0.078}_{-0.080}\ M_J$ | P. Cortés-Zuleta et al. (2020) |
| $R_p$ | $1.415^{+0.044}_{-0.048}\ R_J$ | P. Cortés-Zuleta et al. (2020) |
| Period | $0.78883852^{+75}_{-82}$ days[a] | P. Cortés-Zuleta et al. (2020) |
| $T_{\text{eq}}$ | $2113 \pm 29$ K | P. Cortés-Zuleta et al. (2020) |
| $a$ | $0.01652^{+0.00050}_{-0.00056}$ au | P. Cortés-Zuleta et al. (2020) |
| $a/R_*$ | $3.533^{+0.048}_{-0.052}$ | P. Cortés-Zuleta et al. (2020) |
| $b$ | $0.667^{+0.009}_{-0.009}$ | P. Cortés-Zuleta et al. (2020) |
| Inclination | $79.08^{+0.34}_{-0.37}$ deg | P. Cortés-Zuleta et al. (2020) |
| Eccentricity | $0.0126^{+0.0140}_{-0.0089}$ | P. Cortés-Zuleta et al. (2020) |

**Note.**
[a] The reported uncertainties correspond to the last two digits in the parameter value.

T. Mikal-Evans et al. (2019) This Gaussian process kernel has four free parameters: the covariance amplitude $A$ and the correlation length scales $L_\phi$, $L_y$, and $L_x$. In addition, the model parameterizes the white noise with a rescaling factor $\beta$ for the formal photon noise. To correct for visit-long trends, we tested both linear and quadratic functions in time. We additionally tested fits where we trimmed the second orbit from the five-orbit-long visits and found that this had no significant effect on the results. We adopted uniform priors for the eclipse, baseline, and exponential model parameters. In the Gaussian process model, we followed the approach of N. P. Gibson et al. (2017) and T. M. Evans et al. (2018). We adopted Gamma priors of the form $p(A) \propto e^{-100}A$ for the covariance amplitude, $A$. This favors smaller amplitudes and prevents a small number of outliers disproportionately increasing the amplitude. For the covariance length scales, we fit for the natural log of the inverse correlation length, and applied a uniform prior on each. This favors longer correlation length scales in order to capture the lower-frequency systematics in the data like the orbit-long "hook" shapes. We found that different prior bounds had little effect on the fit results.

We fit these models using the `emcee` software package (D. Foreman-Mackey et al. 2013) to perform a Markov Chain Monte Carlo (MCMC) analysis. When we compared fits using our two instrumental noise models we found that our choice of model did not significantly affect the calculated white-light eclipse depth or its uncertainty, with the values from both fits falling within $1\sigma$ of each other. We elect to use the results from the Gaussian process model in our subsequent analysis, as this fit yielded both the smallest standard deviation of the residuals and the smallest error bars on the fitted model eclipse depth parameter. The results of the white-light light-curve fits are given in Table 2, and the raw white-light light curves with the fitted transit models are shown in Figure 1.

### 3.2. Secondary-eclipse Spectroscopic Light Curves

We constructed spectroscopic light curves by splitting the spectra into six wavelength bins, each with a width of

Table 2
Priors and Posteriors for the White-light Light-curve Fits

| Parameter | Prior | Posterior | |
|---|---|---|---|
| | | Secondary Eclipse | Primary Transit[a] |
| $F_p/F_*$ | $\mathcal{U}(0, 0.5)$ | $0.00142 \pm 0.00008$ | ... |
| Radius ratio $R_p/R_*$ | $\mathcal{U}(0, 0.5)$ | ... | $0.14492^{+0.00076}_{-0.00076}$ |
| Transit center time $t_0$ [BJD$_{UTC}$] | $\mathcal{U}(-0.004, 0.004)$[b] | $2456819.613679^{+561}_{-628}$[c] | $2456820.796002^{+89}_{-81}$[c] |
| Linear slope $m$ | $\mathcal{U}(-0.1, 0.1)$ | $-0.00472^{+0.00082}_{-0.00065}$ | $-0.00532^{+0.00190}_{-0.00209}$ |
| Constant $c$ | $\mathcal{U}(0.9, 1.1)$ | $0.99817^{+0.00030}_{-0.00080}$ | $1.00004^{+0.00040}_{-0.00066}$ |
| Covariance amplitude $A$ | $\Gamma(-1, 0.01)$ | $0.00054^{+0.00035}_{-0.00082}$ | $0.00088^{+0.00036}_{-0.00052}$ |
| Log-inverse covariance length $\ln L_\phi^{-1}$ | $\mathcal{U}(-8, 8)$ | $-2.33864^{+1.46040}_{-1.77957}$ | $-3.34473^{+0.95647}_{-1.35753}$ |
| Log-inverse covariance length $\ln L_y^{-1}$ | $\mathcal{U}(-8, 8)$ | $-4.44326^{+3.49969}_{-2.90728}$ | $-2.02523^{+0.81901}_{-0.83972}$ |
| Log-inverse covariance length $\ln L_x^{-1}$ | $\mathcal{U}(-12, 8)$ | $2.05124^{+5.84237}_{-3.60598}$ | $-1.63768^{+0.67831}_{-0.69313}$ |
| White noise parameter $\beta$ | $\mathcal{U}(0.1, 10)$ | $1.23593^{+0.14090}_{-0.12468}$ | $1.29472^{+0.10931}_{-0.10946}$ |

**Notes.**
[a] We fit the transit using a four-parameter nonlinear limb-darkening law with coefficients $[u_1, u_2, u_3, u_4] = [0.6087, 0.0177, -0.0329, -0.0204]$.
[b] The priors on the secondary-eclipse and primary-transit mid-times are centered at 2456819.614610 and 2456820.79603 BJD, respectively ($t_0 = 0$).
[c] The reported uncertainties correspond to the last three digits and two digits in the parameter value, respectively.

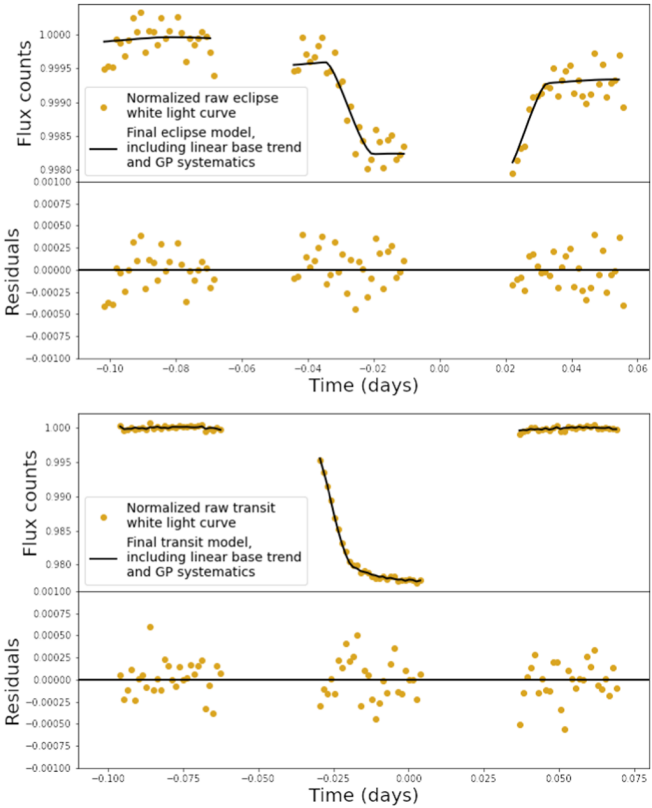

**Figure 1.** White-light light curves for the secondary eclipse (top) and primary transit (bottom). The fitted model is shown in the black lines, and includes the transit model, linear base trend, and GP systematics.

$\Delta\lambda \sim 0.08$ μm, or approximately 17 pixels. Like C. M. Huitson et al. (2013), we found that this number of bins was sufficient to spectroscopically resolve the water feature while still ensuring that there is enough flux in each wavelength bin to fit for the systematic trends. We then employed a "common-mode" correction to remove wavelength-independent systematic trends in the raw light curves (T. Mikal-Evans et al. 2019; J. J. Spake et al. 2021). This process entails correcting the systematic trends common to all wavelengths by dividing the white-light light curve by its best-fit eclipse model, and then dividing the result from each spectroscopic light curve before the spectroscopic fitting procedure.

For the spectroscopic light-curve fitting, we fixed the time of mid-transit $t_0$ to the best-fit value from the MCMC analysis for the white-light light curve. Otherwise, we used a similar procedure to the white-light light-curve fitting, again adopting Gaussian priors (GPs) to fit the HST systematics and letting the eclipse depth vary as a free parameter. Refitting the GP systematics model in each spectroscopic bandpass accounts for the wavelength-dependent component in the systematics, which would not have been resolved by only the "common-mode" corrections. These wavelength-dependent components can be attributed to the fact that the timescale of the charge trapping is proportional to the rate of photons hitting the detector, which varies with wavelength and depends on the shape of the stellar spectrum. The resulting eclipse depths are given in Table 3, and the spectroscopic light curves are displayed in Figure 2.

### 3.3. Primary-transit Light Curves

We also calculated a transmission spectrum for WASP-19b using the data from visit 2, which covered a primary transit. We followed the same procedures outlined above. For the primary-transit model, we allowed two transit shape parameters ($R_P/R^*$ and $t_0$) to vary in the white-light light-curve fit, and then fixed $t_0$ in the fits to individual bandpasses. To obtain the limb-darkening parameters in each bandpass, we used the four-parameter nonlinear limb-darkening law (A. Claret 2000) and fitted it to an ATLAS stellar model (F. Castelli & R. L. Kurucz 2003) that corresponded to the star WASP-19 (with $\log(g) = 4.5$ and $T_{eff} = 5500$ K). The results for the transit white-light light-curve fit are also given in Table 2.

Similar to the emission spectrum measurements, we measured the planet's transmission spectrum by splitting the flux into six wavelength channels, with the same instrumental noise models that we used in the white-light light-curve fits. The measured transit depths and the associated limb-darkening coefficients are shown in Table 3.

**Table 3**
Wavelength-dependent Secondary-eclipse and Transit Depths for the Spectroscopic Light Curves

| Secondary-eclipse Measurements | |
|---|---|
| Wavelengths ($\mu$m) | $F_p/F_*$ |
| 1.1339–1.2164 | $0.00109^{+0.00018}_{-0.00016}$ |
| 1.2164–1.2990 | $0.00143^{+0.00025}_{-0.00027}$ |
| 1.2990–1.3861 | $0.00133^{+0.00015}_{-0.00012}$ |
| 1.3861–1.4686 | $0.00124^{+0.00018}_{-0.00016}$ |
| 1.4686–1.5512 | $0.00174^{+0.00018}_{-0.00018}$ |
| 1.5512–1.6383 | $0.00175^{+0.00017}_{-0.00016}$ |
| 3.4125–3.7875 (a) | $0.00485^{+0.00024}_{-0.00024}$ |
| 4.2463–4.7538 (a) | $0.00584^{+0.00029}_{-0.00029}$ |
| 5.4500–6.1500 (b) | $0.00650^{+0.00011}_{-0.00011}$ |
| 7.2750–8.7250 (b) | $0.00730^{+0.00012}_{-0.00012}$ |

| Primary-transit Measurements | |
|---|---|
| Wavelengths ($\mu$m) | $R_p/R_*$ |
| 1.1331–1.2157 | $0.14458^{+0.00076}_{-0.00077}$ |
| 1.2157–1.2982 | $0.14424^{+0.00051}_{-0.00056}$ |
| 1.2982–1.3853 | $0.14471^{+0.00102}_{-0.00060}$ |
| 1.3853–1.4679 | $0.14599^{+0.00082}_{-0.00089}$ |
| 1.4679–1.5504 | $0.14660^{+0.00065}_{-0.00075}$ |
| 1.5504–1.6375 | $0.14463^{+0.00101}_{-0.00063}$ |

| Primary-transit Limb-darkening Coefficients | |
|---|---|
| Wavelengths ($\mu$m) | $[u_1, u_2, u_3, u_4]$ |
| 1.1331–1.2157 | [0.6421, −0.3373, 0.5591, −0.2770] |
| 1.2157–1.2982 | [0.5985, −0.1418, 0.2992, −0.1772] |
| 1.2982–1.3853 | [0.5690, 0.0286, 0.0448, −0.0688] |
| 1.3853–1.4679 | [0.5602, 0.1872, −0.2343, 0.0563] |
| 1.4679–1.5504 | [0.6005, 0.2083, −0.3733, 0.1314] |
| 1.5504–1.6375 | [0.6920, 0.0853, −0.3737, 0.1620] |

**Note.** We included the published Spitzer secondary-eclipse measurements from (a) D. R. Anderson et al. (2013) and (b) I. Wong et al. (2016) in this table as these were used in our atmospheric retrievals.

## 4. Atmosphere Models

The results for the emission and transmission spectra are displayed in Figure 4. We initially focus on the emission spectroscopy data in our retrievals, as the high stellar activity level causes the measured shape of the transmission spectrum to vary from epoch to epoch (see discussion in Section 5.1). In order to obtain the best possible constraints on the planet's atmospheric properties in our retrievals, we also included published Spitzer secondary-eclipse photometry measured from D. R. Anderson et al. (2013) and I. Wong et al. (2016) in the 3.6, 4.5, 5.8, and 8.0 $\mu$m bands. I. Wong et al. (2016) also published transit photometry, but we have chosen not to include them because of potential biases due to stellar variability. Section 5.1.1 will discuss this further and calculate the expected variability amplitude in the Spitzer bandpasses. Although J. L. Bean et al. (2013) also measured a spectroscopically resolved near-infrared emission spectrum spanning a similar wavelength range to ours, we found that our results disagreed with their ground-based measurements in the overlapping wavelengths; this may be due to the more complex systematic noise models needed to account for time-varying atmospheric and instrumental conditions in ground-based observations, which typically exhibit higher levels of correlated noise than space-based photometry. We therefore elected to exclude the measurements from J. L. Bean et al. (2013) from our retrievals, and instead jointly fit the HST WFC3/G141 and Spitzer photometry.

### 4.1. Atmospheric Retrieval

We performed an atmospheric retrieval using the PLATON package (M. Zhang et al. 2019, 2020). PLATON is a forward modeling and retrieval tool that can fit both transmission and emission spectroscopy, calculating the abundances and line opacities of 28 molecules. When fitting the dayside emission spectrum, we used a cloud-free atmosphere, allowed the atmospheric metallicity and C/O ratio to vary as free parameters in our equilibrium chemistry fits, and retrieved a parametric $T$–$P$ profile as described in M. R. Line et al. (2013). This $T$–$P$ profile has five free parameters: the thermal opacity $\kappa_{\rm th}$, the visible-to-thermal opacity ratio of the first visible stream $\gamma$, the visible-to-thermal opacity ratio of the second visible stream $\gamma_2$, the percentage apportioned to the second visible stream $\alpha$, and the effective albedo $\beta$. We also ran joint retrievals including our HST transmission spectrum. In these fits, we added the planetary radius as a free parameter and allowed for a gray opaque cloud deck where the cloud-top pressure is a free parameter in the fit. For the joint fits including the transmission retrieval, we assume that the atmosphere in the terminator region can be reasonably well approximated by an isothermal model, and retrieve for the temperature at the limb as an additional free parameter.

We began by carrying out a combined retrieval on our measured HST transmission and emission spectrum to evaluate the strength of the water detection in the 1.1–1.6 $\mu$m wavelength range. We fit the atmosphere model to the data using the dynesty nested sampling package (S. Koposov et al. 2023) with 1000 live points and a resolution of 10,000. The resulting best-fit models are shown in Figure 4. We find that there is a dip in the emission spectrum and a bump in the transmission spectrum at 1.4 $\mu$m, corresponding to water absorption. We show the resulting best-fit parametric dayside $T$–$P$ profile in Figure 3 and list the retrieved atmospheric parameters in Table 4.

We measured a detection significance of 2.1$\sigma$ for the water feature in emission, and 2.3$\sigma$ for the water feature in transmission. This is lower than B. Edwards et al.'s (2023) measured water significance of 4$\sigma$ in transmission, which was based on a combined reduction of both WFC3 transit observations. To calculate this value, we fitted a blackbody model (i.e., with no water opacity), calculated the ratio of the Bayesian evidences from the blackbody and the retrieved model with water, and converted to a significance level. For the transmission spectrum, unlike B. Edwards et al. (2023), who used a flat line as their water-free model, we ran the fiducial model with zero water opacity to represent the water-free model, and obtained the Bayesian factor therewith. This difference in method would therewith affect the resulting calculated significances.

We find that although the detection of water absorption rules out high C/O ratios, our fits otherwise give relatively weak constraints on the atmospheric C/O ratio. This is expected, as the 1.1–1.6 $\mu$m wavelength range of the HST WFC3 data does not contain any strong absorption features from CO and $CO_2$, which are expected to be the major carbon-bearing species in

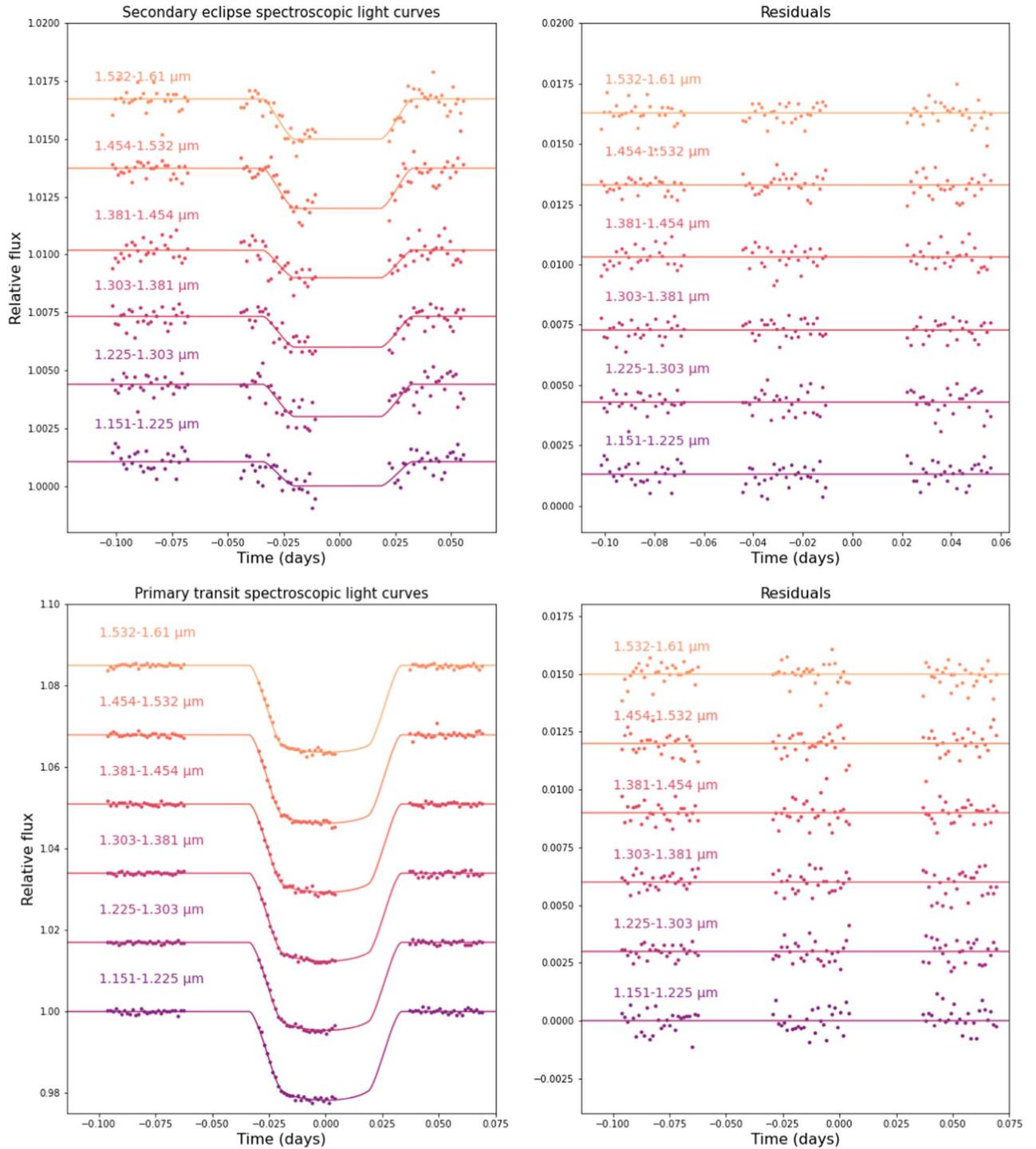

**Figure 2.** Spectroscopic light curves for the secondary eclipse (top) and primary transit (bottom), spanning the 1.1–1.6 $\mu$m wavelength range. Left panel: each light curve has been divided by the best-fit systematic model; overplotted are the best-fit eclipse models as solid lines. Right panel: fit residuals for each of the spectroscopic light curves. We applied vertical offsets in both panels for clarity.

WASP-19b's atmosphere. We therefore expand our fits to include the published photometric Spitzer secondary-eclipse points, which extend to longer wavelengths and increase our sensitivity to carbon-bearing molecules. We carry out an emission-only retrieval with the Spitzer data points and a combined retrieval including our HST transmission spectrum in order to evaluate the relative influence that our HST transmission spectrum has on the results. As shown in Figure 4, adding the Spitzer data points modestly improved our constraints on the atmospheric C/O ratio, while leaving the retrieved metallicity effectively unconstrained. Overall, we find that we obtain the tightest constraints on the atmospheric composition when we also include our HST transmission spectrum, although the improvement is incremental when compared to the dayside-only fits including the Spitzer photometry. We show a corner plot with the posterior distributions for the combined retrieval with both Spitzer and HST data in Figure 5.

Table 4
Table of Atmospheric Retrieval Priors and Posteriors

| Parameter | Prior | Dayside Retrieval, HST + Spitzer | | Combined Retrieval, only HST | | Combined Retrieval, HST + Spitzer | |
| --- | --- | --- | --- | --- | --- | --- | --- |
| | | Best Fit | Posterior (Median $^{+1\sigma}_{-1\sigma}$) | Best Fit | Posterior (Median $^{+1\sigma}_{-1\sigma}$) | Best Fit | Posterior (Median $^{+1\sigma}_{-1\sigma}$) |
| log (Z) | $\mathcal{U}(-1, 3)$ | 0.53 | $0.56^{+1.69}_{-1.18}$ | −0.22 | $0.89^{+1.31}_{-1.25}$ | −0.93 | $0.66^{+1.45}_{-1.10}$ |
| C/O ratio | $\mathcal{U}(0.2, 2)$ | 0.89 | $0.91^{+0.71}_{-0.45}$ | 0.79 | $0.87^{+0.74}_{-0.44}$ | 0.68 | $0.76^{+0.76}_{-0.38}$ |
| log ($P_{\text{cloudtop}}$) | $\mathcal{U}(-1, 5)$ | ⋯ | ⋯ | 4.52 | $2.72^{+1.57}_{-2.25}$ | 4.88 | $3.10^{+1.31}_{-2.27}$ |
| $T_{\text{limb}}$ [K] | $\mathcal{U}(1500, 3000)$ | 2671 | $2316^{+457}_{-523}$ | 2412 | $2169^{+521}_{-450}$ | 2032 | $2226^{+474}_{-463}$ |
| $R_P$ [$R_{\text{Jup}}$] | $\mathcal{U}(1.27, 1.55)$ | ⋯ | ⋯ | 1.38 | $1.38^{+0.03}_{-0.04}$ | 1.40 | $1.38^{+0.03}_{-0.03}$ |
| log ($k_{\text{th}}$) | $\mathcal{U}(-5, 1)$ | −1.44 | $-2.31^{+0.94}_{-1.76}$ | −2.37 | $-2.87^{+1.31}_{-1.45}$ | −1.98 | $-2.55^{+1.08}_{-1.70}$ |
| log ($\gamma$) | $\mathcal{U}(-4, 1)$ | −0.79 | $-0.17^{+0.59}_{-0.42}$ | −0.34 | $-0.06^{+0.52}_{-0.45}$ | −0.78 | $-0.12^{+0.50}_{-0.37}$ |
| log ($\gamma_2$) | $\mathcal{U}(-4, 2)$ | −0.34 | $-0.06^{+0.70}_{-0.65}$ | −0.06 | $-0.1^{+0.68}_{-0.58}$ | 0.52 | $-0.04^{+0.63}_{-0.58}$ |
| $\alpha$ | $\mathcal{U}(0, 0.5)$ | 0.18 | $0.25^{+0.18}_{-0.16}$ | 0.37 | $0.25^{+0.17}_{-0.17}$ | 0.08 | $0.25^{+0.17}_{-0.17}$ |
| $\beta$ | $\mathcal{U}(0.5, 2.5)$ | 0.88 | $1.07^{+0.11}_{-0.17}$ | 1.11 | $1.12^{+0.09}_{-0.24}$ | 0.89 | $1.12^{+0.08}_{-0.17}$ |

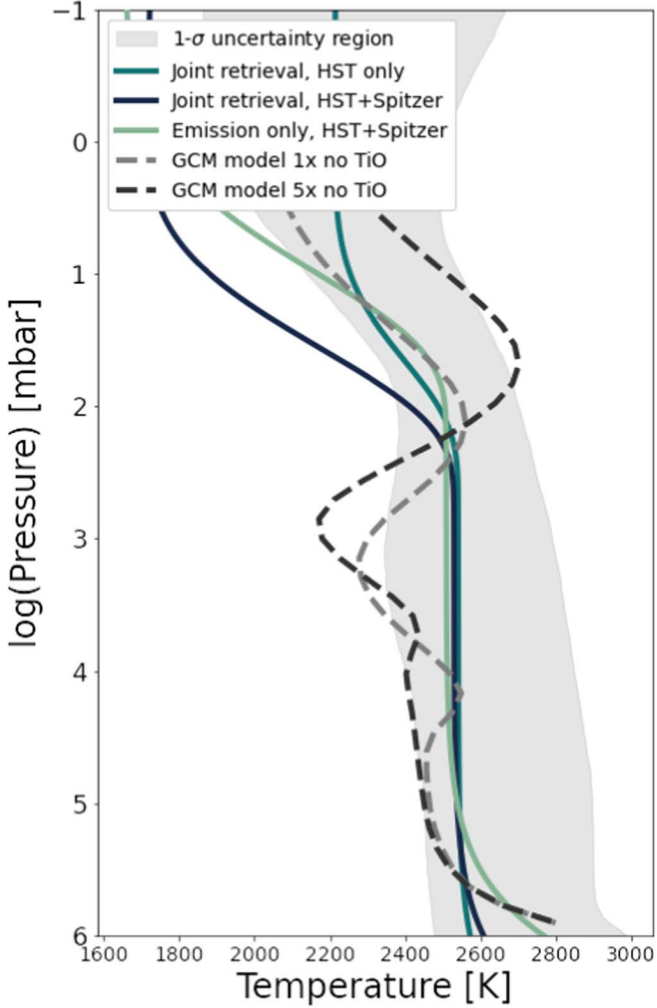

**Figure 3.** Dayside T–P profiles for our three best-fit models, as well as two of the GCM models from T. Kataria et al. (2016). The best-fit T–P profile from the combined retrieval (including the HST and Spitzer emission spectra and the HST transmission spectrum) is in solid black, along with its 1σ uncertainty region shown as the shaded gray region. We also include the best-fit T–P profiles from the combined retrieval with HST transmission and emission spectroscopy data only and the dayside-only retrieval with HST and Spitzer data. The GCM model with 1× and 5× solar metallicity, both without TiO, are displayed by the dashed lines.

### 4.2. Comparison to Predictions from a General Circulation Model

T. Kataria et al. (2016) published GCMs for a sample of nine hot Jupiters including WASP-19b; a more extensive set of GCM models for WASP-19b is also described in I. Wong et al. (2016). These models assume that the atmospheric chemistry is in local thermal equilibrium with the exception of TiO and VO, whose abundances are set to zero. These two molecules condense at relatively high temperatures and are expected to be removed from the atmosphere by cold traps on the planet's nightside (D. S. Spiegel et al. 2009; V. Parmentier et al. 2013). Because they have broad absorption features at optical wavelengths, these two molecules are thought to be the primary drivers of thermal inversions in hot Jupiter atmospheres (I. Hubeny et al. 2003). The fact that WASP-19b has water absorption in its dayside emission spectrum provides additional confirmation that these gases are not present at the predicted equilibrium abundances, since the presence of these gases usually indicates a thermal inversion, which we do not see from our retrievals. This result confirms the findings from Q. Changeat et al. (2022), who also reported a noninverted temperature profile most likely due to the lack of metal oxide gases. Figure 3 shows the dayside-averaged temperature profiles predicted by the GCM models, which follow a broadly similar shape. Although the GCM profiles contain several weak inversions, their integrated dayside emission spectra in Figure 4 show water absorption features with an amplitude comparable to the size of our measured water absorption feature. This suggests that our observations are primarily probing layers in the GCM models where temperature is decreasing with increasing height, consistent with our simpler parametric T–P profile retrieval results.

Figure 4 shows that both our 1.1–1.6 μm emission spectrum and the Spitzer photometry at longer wavelengths are in good agreement with the prediction from two GCM models, one at solar metallicity and the other with 5× solar metallicity. Both models assume a solar C/O ratio and both have gas-phase TiO and VO removed. Our retrieved dayside emission spectrum lies between the 1× and 5× solar metallicity models. These same models also provide a good match to the 3.6 and 4.5 μm Spitzer phase curves for WASP-19b reported in I. Wong et al. (2016). These models do not include opacity from condensate clouds, nor do they account for the potential effects of clouds on the atmospheric temperature structure

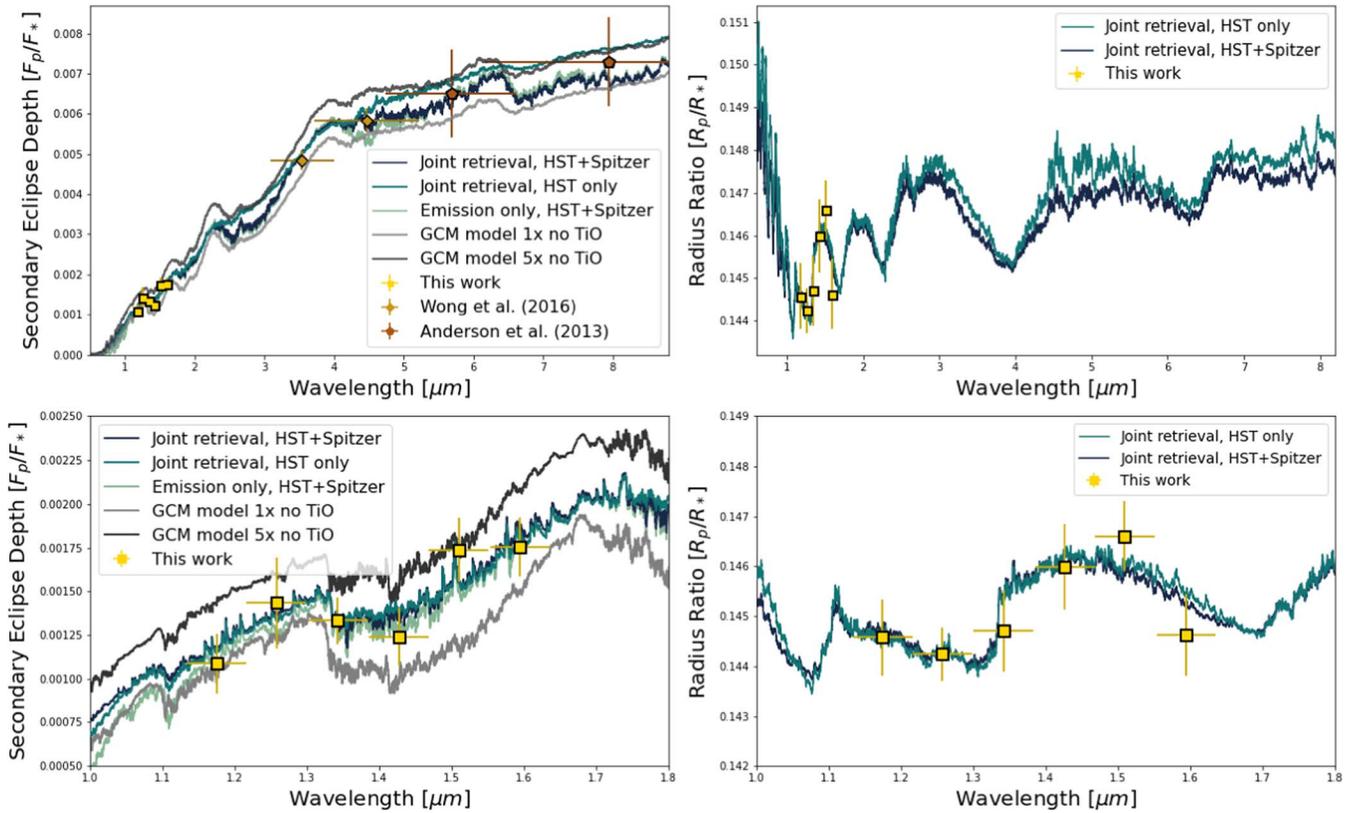

**Figure 4.** Retrieved emission (top left) and transmission (top right) spectra spanning 0.8–8.3 μm, with three best-fit models corresponding to a combined retrieval with our HST transmission and emission data only, a combined retrieval with both HST and Spitzer data, and a dayside-only retrieval with HST and Spitzer data. The lower half shows a closer view of the retrieved emission and transmission spectra in the 1.1–1.8 μm range of the HST data. The data used in the retrievals are shown by the colored markers, as displayed in the legends. Also included are the GCM models with 1× (gray line) and 5× (black line) solar metallicities with no TiO, from T. Kataria et al. (2016).

(M. Roman & E. Rauscher 2019; V. Parmentier et al. 2020; M. T. Roman et al. 2021). The relatively good agreement with our measured dayside emission spectrum suggests that clouds have a minimal effect on WASP-19b's dayside atmosphere. This is consistent with model predictions, which find that WASP-19b's dayside is likely too hot for most condensates (C. M. Huitson et al. 2013).

The GCM models presented here neglect potential magnetic drag effects, which can act as a drag force on the zonal winds (R. Perna et al. 2010; K. Menou 2011). For more highly ionized atmospheres, these magnetic forces can even reverse the atmospheric flows, resulting in a dayside hot spot that is shifted to the west rather than the east (A. W. Hindle et al. 2021a, 2021b). The relatively good agreement between our measured flux levels and the dayside emission spectrum predicted by these GCMs indicates that magnetic drag effects are relatively unimportant in WASP-19b's atmosphere. A. W. Hindle et al. (2021b) found that for HAT-P-7b and Kepler-76b, which have dayside temperatures similar to that of WASP-19b, magnetic effects can produce a hot-spot reversal for magnetic field strengths greater than 5−10 G. However, I. Wong et al. (2016) find that this planet's phase curve is consistent with a hot-spot offset of zero. This suggests that either WASP-19b's dipole magnetic field strength is weaker than 5−10 G or that its atmosphere is more weakly ionized than predicted by these GCM models, perhaps as a result of a significantly substellar abundance of alkali metals, which are the primary source of ions in hot Jupiter atmospheres (K. Batygin & D. J. Stevenson 2010; R. Perna et al. 2010).

## 5. Discussion

### 5.1. WASP-19b's Transmission Spectrum in Context

There are multiple published transmission spectra for WASP-19b in both the optical and infrared, and in principle we could improve our constraints on this planet's cloud properties and atmospheric composition if we included these data sets in our retrievals. In this section, we compare these spectra to our measured HST transmission spectrum, and discuss the limitations that this star's relatively high activity level imposes on our ability to combine transmission spectroscopy data from multiple epochs. We begin by comparing our HST WFC3 transmission spectrum to published observations of WASP-19b obtained using the same instrument, taken at multiple epochs, and then expand our discussion to include transmission spectra obtained from the ground at optical wavelengths, which we expect to be even more strongly affected by stellar activity.

#### 5.1.1. HST WFC3 Transmission Spectra

As discussed in Section 1, the WFC3 transmission spectrum presented in C. M. Huitson et al. (2013) utilized the 2011 transit observation, while B. Edwards et al. (2023) combined this data set with the 2014 transit observation to create a composite transmission spectrum. Like Q. Changeat et al. (2022), we only analyzed the 2014 transit observation. Both our study and the C. M. Huitson et al. (2013) study extracted the data using six wavelength bins, while B. Edwards et al.

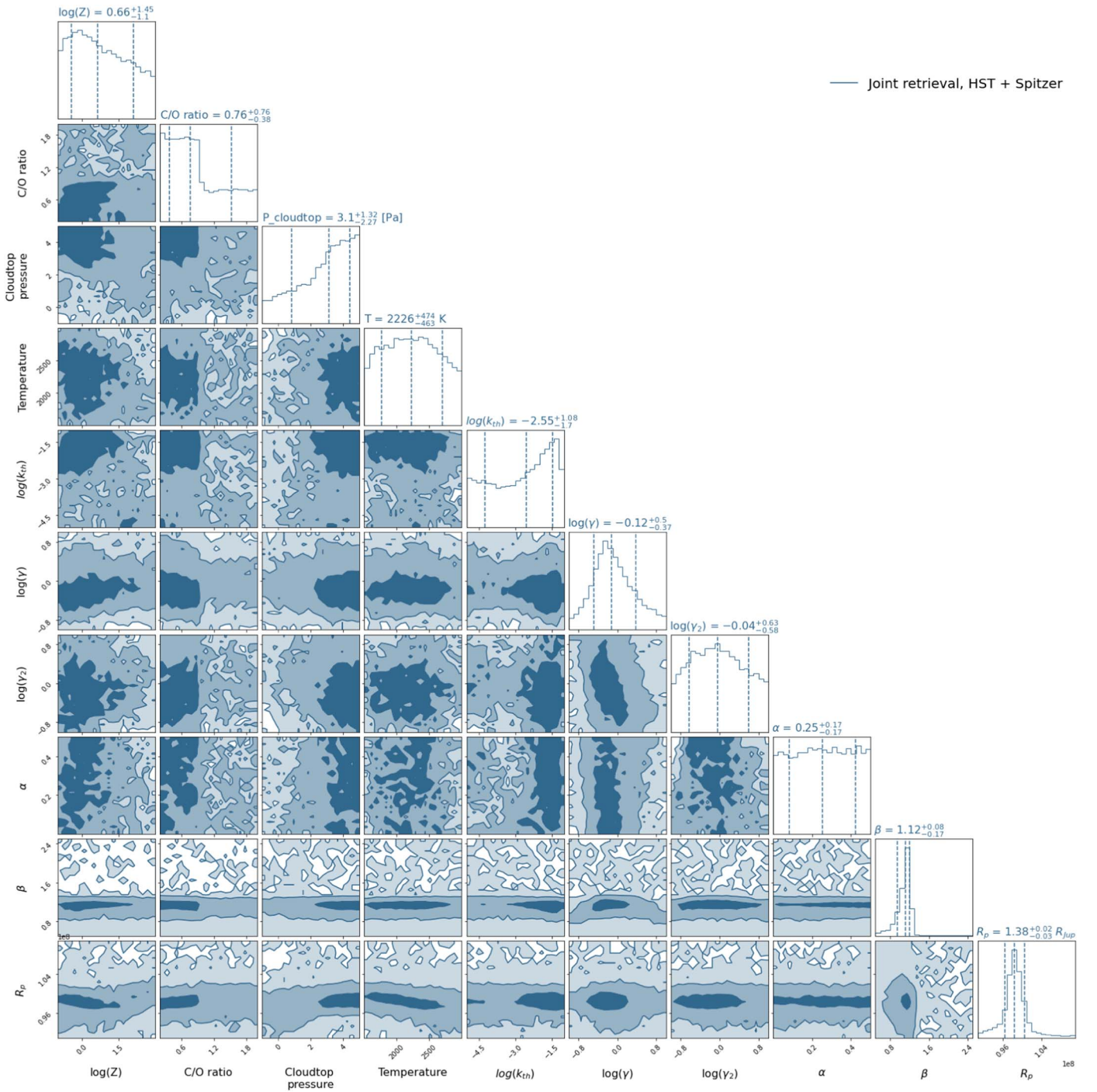

**Figure 5.** Corner plot displaying the posterior probability distributions for our combined retrieval with both HST and Spitzer data points.

(2023) and Q. Changeat et al. (2022) extracted their transmission spectra at a higher resolution. We plot these published transmission spectra and compare them to ours in Figure 6. We find that the overall shapes of the four transmission spectra are in good agreement, but their average values are positively offset by approximately 0.0041 (C. M. Huitson et al. 2013) and 0.0021 (B. Edwards et al. 2023) in $R_p/R_*$. We find that the Q. Changeat et al. (2022) transmission spectrum, which is also based solely on the 2014 transit observation, largely overlaps with ours despite using the same analysis framework as B. Edwards et al. (2023). This suggests that it is the 2011 transit data which are pulling the joint result in B. Edwards et al. (2023) to lower radius ratios.

It is tempting to attribute these offsets to stellar activity, as the two transit observations were taken 3 yr apart. However, the offset between our measured transmission spectrum using the 2014 data and the transmission spectrum measured by C. M. Huitson et al. (2013) using the 2011 data is larger than the expected maximum offset caused by stellar variability. Published studies indicate that WASP-19's optical flux can vary in brightness by up to 3% (I. Wong et al. 2020). Following B. V. Rackham et al. (2018), we used this percentage flux difference to estimate the resulting change in transit depth due to stellar variability, assuming that the variability is predominantly due to starspots:

$$D_{\lambda,\text{obs}} = \frac{D_\lambda}{1 + f_{\text{spot}}\left(1 + \frac{F_{\text{spot}}}{F_{\text{star}}}\right)}, \quad (2)$$

where $D_\lambda$ is the true transit depth, $D_{\lambda,\text{obs}}$ is the observed transit depth with the effects of stellar variability, $f_{\text{spot}}$ is the fraction of spot coverage, and $F_{\text{spot}}$ and $F_{\text{star}}$ are the averaged spectra of a starspot and star, respectively. We used the PHOENIX catalog of stellar spectra to obtain a spectrum closest to that of WASP-19 ($T_{\text{eff}} = 5500$ K, $\log(g) = 4.5$), and assumed the starspots have a $T_{\text{eff}}$ approximately 500 K cooler than the stellar photosphere (C. M. Huitson et al. 2013). We took the ratio of the averaged fluxes over the optical wavelength range (in increments of 0.1 $\mu$m), and used the maximum stellar flux variability factor of 0.03 to estimate the range of spot coverage fractions over the wavelength range. We then used the above equation to rescale our transmission spectrum and found that this variability can produce offsets of up to 0.001 in $R_p/R_*$ in the WFC3 bandpass, and up to 0.00054 and 0.00055 in the 3.6 $\mu$m and 4.5 $\mu$m Spitzer bandpasses, respectively (see Figure 6). We conclude that starspots alone cannot explain the discrepancy between our 2014 transmission spectrum and the 2011 spectrum from C. M. Huitson et al. (2013).

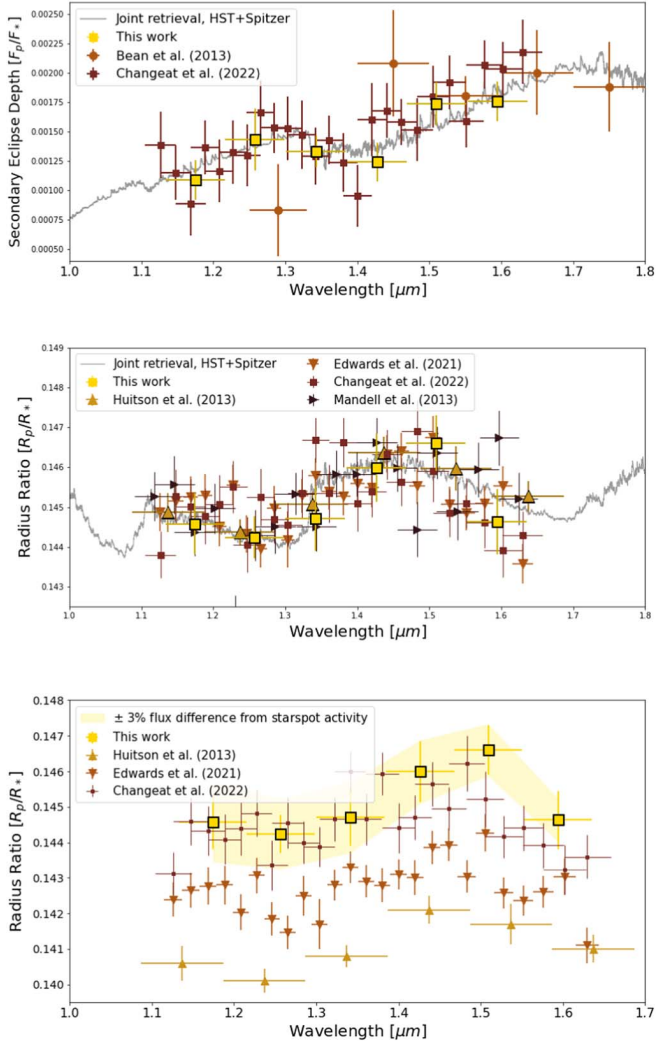

**Figure 6.** Top: our HST WFC3 emission spectrum compared to the one published by J. L. Bean et al. (2013) and Q. Changeat et al. (2022). Middle: our transmission spectrum from this study compared to the one published in A. M. Mandell et al. (2013), C. M. Huitson et al. (2013), B. Edwards et al. (2023), and Q. Changeat et al. (2022), with offsets added to their flux (see Figure 7 for details on the offsets). Bottom: our HST WFC3 transmission spectrum with the allowed range of flux offsets due to a 3% stellar variability, compared with other published transmission spectra by C. M. Huitson et al. (2013) and B. Edwards et al. (2023), displaying the flux differences.

Next, we confirmed that our fixed values for the inclination, $a/R_{\text{star}}$, and impact parameter are consistent with the values reported in C. M. Huitson et al. (2013), making differences in these values unlikely to explain the offsets. We also considered differences in limb-darkening models. WASP-19b has a relatively high impact parameter ($b \approx 0.667$), and it is therefore more sensitive to our choice of limb-darkening models than most hot Jupiters. We used a four-parameter limb-darkening model in our fits while C. M. Huitson et al. (2013) used a three-parameter model. Lastly, it is worth noting that the 2011 transit observation was obtained in stare mode while the 2014 observation was obtained in scan mode. Previous studies comparing these two modes for the same target (L. Kreidberg et al. 2014a) have found that the stare-mode observations are generally of lower quality than the scan-mode observations. Furthermore, Figure 3 of C. M. Huitson et al. (2013) displays a raw white-light light curve for the STIS transit with stronger instrumental effects and a more complex structure than in the HST observations we use, which may bias the white-light transit depth measurements.

Ultimately, we attribute these offsets to a combination of differing instrumental effects between the two visits, limb-darkening and impact parameter differences in the fits, and stellar activity. Nevertheless, across these different epochs and different reductions, we find that the shape of the water absorption feature in transmission remains robust and stable.

### 5.1.2. Constraints on Atmospheric Composition from HST Transmission Spectroscopy

The presence of a water feature in the transmission spectrum indicates that the atmospheric C/O ratio must be less than ∼0.9 (N. Madhusudhan et al. 2011; J. I. Moses et al. 2013), but its interpretation is degenerate with multiple other atmospheric parameters given the relatively narrow wavelength coverage of these data. Despite our weak constraints on atmospheric composition, our retrieved values for both the C/O ratio and the metallicity are consistent with those from the retrievals of B. Edwards et al. (2023) within their 1$\sigma$ error bars. Our combined transmission-and-emission retrieval using the Spitzer eclipse measurements gave a C/O ratio of $0.76^{+0.76}_{-0.38}$, and $\log(Z) = 0.66^{+1.45}_{-1.10}$, while B. Edwards et al. (2023) found $0.47^{+0.26}_{-0.28}$ and $0.95^{+0.57}_{-1.11}$, respectively. These results echo the results from the earlier study by C. M. Huitson et al. (2013), which placed an upper bound on the atmospheric C/O ratio.

The presence or absence of starspots can also affect the depth of the measured water absorption feature, as discussed in B. V. Rackham et al. (2018). We explored possible degeneracies between spot coverage and water abundance by performing a retrieval where we allowed the spot coverage to vary as a free parameter while fixing the spot temperature to a value of 5000 K (500 K cooler than the stellar effective temperature), and found that our posterior probability distributions for C/O ratio and metallicity were unchanged. This is not surprising, as both parameters fill their prior ranges in our nominal retrieval. C. M. Huitson et al. (2013) also explored the effects of varying starspot modeling choices on their inferred atmospheric composition and found that it was difficult to constrain WASP-19b's atmospheric properties while also accounting for stellar activity without optical transmission spectroscopy.

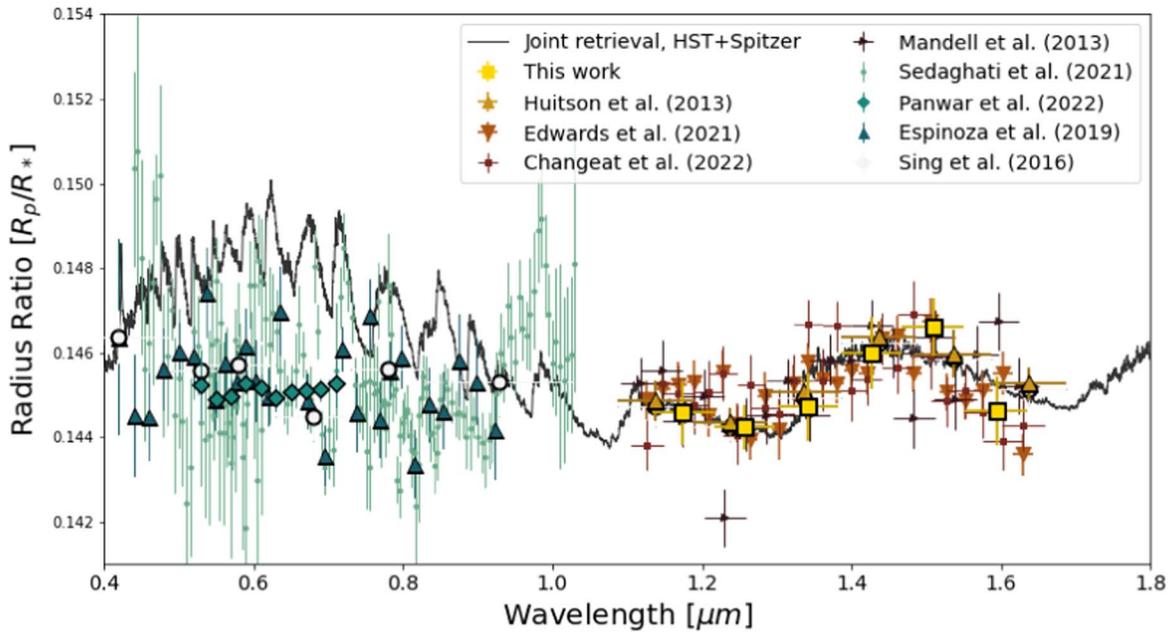

**Figure 7.** Our retrieved transmission spectra from the joint fit to the Spitzer emission photometry and HST transmission and emission spectroscopy projected into the optical range (black line), compared to previously published optical transmission spectroscopy observations by D. K. Sing et al. (2016, with a positive 0.005 offset), E. Sedaghati et al. (2017, positive [0.0031, 0.0065, 0.0058] offsets), N. Espinoza et al. (2019, positive 0.0032 offset), and V. Panwar et al. (2022, negative 0.0016 offset). We also include our HST WFC3 transmission spectrum, along with previously published infrared transmission spectroscopy observations by C. M. Huitson et al. (2013, positive 0.0041 offset), B. Edwards et al. (2023, positive 0.0021 offset), A. M. Mandell et al. (2013, positive 0.0033 offset), and Q. Changeat et al. (2022, positive 0.0005 offset). All the previously published transmission spectra have been normalized to D. K. Sing et al.'s (2016) spectrum.

### 5.1.3. Optical Transmission Spectra

There have been multiple observations of WASP-19b's optical transmission spectrum from both ground- and space-based telescopes, spanning wavelengths from 0.3 to 1.0 μm (D. K. Sing et al. 2016; E. Sedaghati et al. 2017, 2021; N. Espinoza et al. 2019; V. Panwar et al. 2022). To see how our retrieved model compares to these observations, we plotted their data on top of our best-fit transmission spectrum in Figure 7, with positive offsets in $R_p/R_*$ applied to match the average values of each transmission spectrum. Overall, we find that the size of the offsets required is similar to that required for the WFC3 transmission spectra, and is consistent with the star's large measured rotational variability level at optical wavelengths (see Figure 7). We also note that there are significant discrepancies between the shapes of the optical transmission spectra measured in these studies, indicating that some of the ground-based observations may suffer from uncorrected wavelength-dependent correlated noise in their light curves.

While it is tempting to include some of these data sets in our retrieval, this would require us to fit for a unique spot coverage fraction at each epoch. Setting aside the ground-based observations for the moment, this would require the addition of five unique spot coverage fractions for the three STIS transits and two WFC3 transits. Alternatively, we could use ground-based monitoring data to independently constrain the relative spot coverage fraction at each epoch (D. K. Sing et al. 2011). Unfortunately, we are not aware of any ground-based monitoring of WASP-19 spanning the epochs of these data sets. We conclude that without strong priors on the instantaneous stellar activity level at each epoch, incorporating these data into our fits would not provide enough additional information to justify the significantly increased fit complexity required.

### 5.2. WASP-19b's Emission Spectrum in Context

#### 5.2.1. HST WFC3 Emission Spectra

Our HST WFC3 emission spectrum is consistent with the WFC3 emission spectrum reported by Q. Changeat et al. (2022) using the same data as part of their population-level study of hot Jupiter emission spectra, despite minor differences in our reduction methods (see Figure 6). They also combine their WFC3 emission spectrum with published Spitzer secondary-eclipse depths in order to constrain WASP-19b's atmospheric composition. Unlike our fits, they carry out free retrievals where they assume that the abundances of different molecular species remain constant as a function of pressure, measuring a water abundance of $\log(H_2O)$ of $-3.0^{+0.6}_{-1.0}$. Figure 8 shows that our sampled $H_2O$ abundance profiles encompass the retrieved value from Q. Changeat et al. (2022), which has tighter constraints due to them combining two transit observations in their fit.

#### 5.2.2. A Population-level View of WASP-19b's Emission Spectrum

Here we compare WASP-19b's dayside emission spectrum to the ensemble of hot Jupiters that have been observed in emission by HST WFC3. First, we quantified the strength of the water feature following the method used by M. Mansfield et al. (2021). We began by fitting a blackbody to the out-of-band regions of the emission spectrum where there is minimal water opacity (1.22–1.33 $\mu$m and 1.53–1.61 $\mu$m). Then, we used the equation from M. Mansfield et al. (2021) to calculate the strength of the water feature in the water band:

$$S_{H_2O} = \log_{10}\left(\frac{F_{B,\text{in}}}{F_{\text{obs,in}}}\right), \quad (3)$$

where $F_{B,\text{in}}$ is the flux of the fitted blackbody, and $F_{\text{obs,in}}$ is the flux of the observed data. A positive $S_{H_2O}$ corresponds to a water absorption feature, a negative $S_{H_2O}$ means water emission. We calculated the value of this index by coarsely binning our measured emission spectrum into three points, corresponding to the in-water-band region and the two out-of-band regions, and taking the mean of the out-of-band fluxes. This method gives us an $S_{H_2O}$ value of $0.079 \pm 0.072$, indicating a 1$\sigma$ water absorption feature and a dayside temperature of $2494 \pm 69$ K.

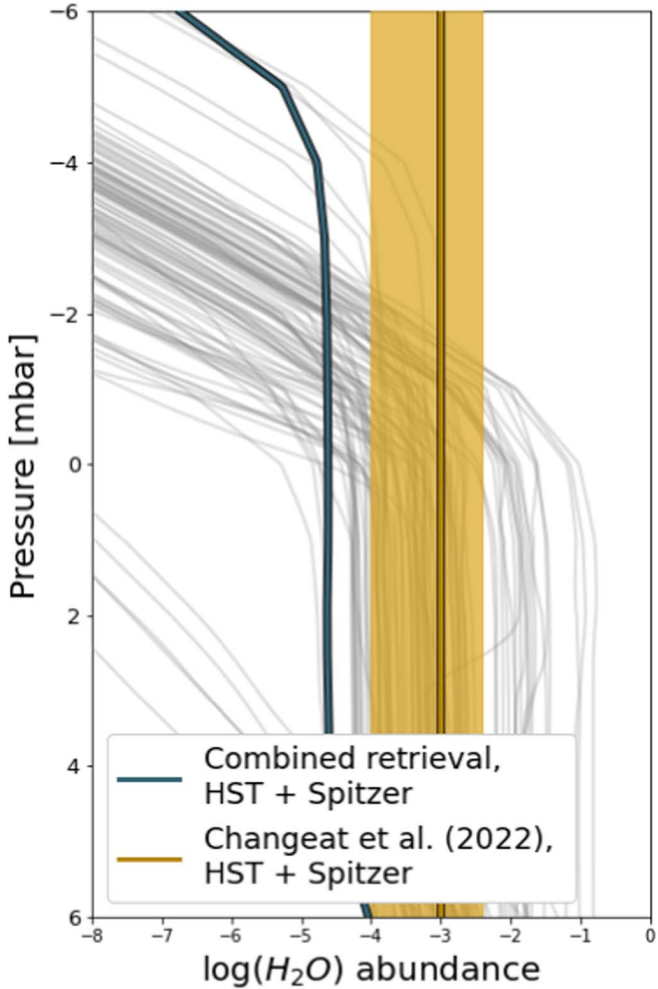

**Figure 8.** Distribution of the $H_2O$ abundance as measured from our best-fit combined retrieval with Spitzer data, compared to the measured water abundance from Q. Changeat et al. (2022). The faded gray lines are randomly drawn samples from our retrieval to illustrate the distribution of possible $H_2O$ abundance profiles.

Next, we consider where our empirically measured water absorption feature strength places WASP-19b in the context of the broader population of hot Jupiters. M. Mansfield et al. (2021) found that as the dayside temperatures of these planets increased they generally exhibited weaker water absorption features. Eventually, these absorption features turn into emission features for planets with dayside temperatures between ~2500 and 3000 K, and for the most highly irradiated planets (dayside temperatures higher than ~3000 K) the feature disappears. This is thought to result from the molecular disassociation of water combined with the addition of continuum opacity from $H^-$ ions, which is a continuum absorber that can act like a cloud to mute the water feature. With a best-fit dayside brightness temperature of $2494 \pm 69$ K outside of the water band in the WFC3 wavelength range and a relatively weak water absorption feature ($S_{H_2O} = 0.079 \pm 0.072$), we find that WASP-19b is located very close to the transition point where water absorption transforms into water emission. Of this sample of planets, WASP-19b appears most similar to HAT-P-41b and WASP-74b, both of which have dayside temperatures between 2300 and 2500 K and water absorption in their dayside emission spectra (G. Fu et al. 2021, 2022). A distribution of $S_{H_2O}$ against $T_{\text{dayside}}$ for a population of hot Jupiters can be seen in Figure 9, following M. Mansfield et al.'s (2021) study.

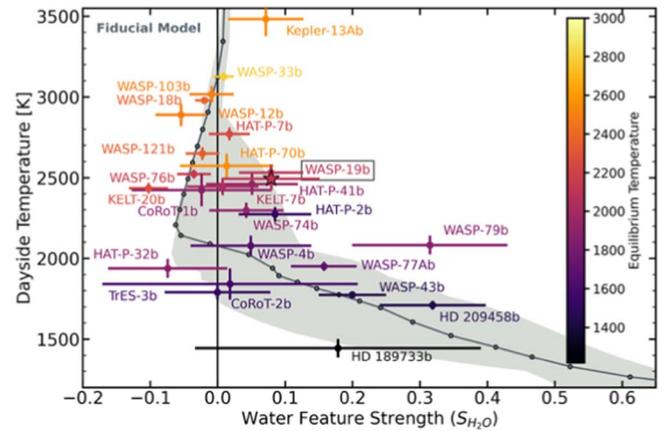

**Figure 9.** Calculated water-feature strength for WASP-19b and its corresponding best-fit dayside temperature (represented by the star) as compared to the broader population of hot Jupiters from M. Mansfield et al. (2021).

## 6. Conclusions

We have presented a reanalysis of a secondary eclipse and transit observation of WASP-19b from the HST's WFC3 spectrograph, spanning a wavelength range of 1.1–1.6 $\mu$m. We fit the spectroscopic light curves from the secondary eclipse and primary transit and found that our transmission and emission spectra are consistent with results from previous studies (C. M. Huitson et al. 2013; Q. Changeat et al. 2022; B. Edwards et al. 2023). We expanded on published analyses by performing atmospheric retrievals on different combinations of data to help constrain its atmospheric parameters, and explored the effect of stellar activity on its transmission spectrum in greater depth. We also compared our results to predictions from GCMs, and compared the size of WASP-19b's dayside water absorption feature to the population of hot Jupiters with similar measurements in the transitional irradiation regime with some emergent temperature inversions (M. Mansfield et al. 2021).

We have presented three atmospheric retrieval models: a dayside-only retrieval, combining HST WFC3 eclipse spectroscopy and Spitzer 3.5–9 $\mu$m eclipse photometry; a retrieval with just HST WFC3 eclipse and transit spectroscopy; and a combined retrieval with both HST WFC3 eclipse and transit spectroscopy plus Spitzer eclipse photometry. All three of these models show a water feature at 1.4 $\mu$m and show no thermal inversion in the T–P profile. Of these three models, the combined retrieval with both HST and Spitzer data provides the best constraints on the retrieved atmospheric parameters.

We found that our dayside emission spectrum is in good agreement with the emission spectrum predicted by 1–5× solar GCMs from T. Kataria et al. (2016). The close agreement between our observations and the GCM model predictions indicates that magnetic fields do not significantly alter WASP-19b's atmospheric circulation, despite its relatively high atmospheric temperature. Using our measured emission spectrum, we calculate a dayside temperature of $2494 \pm 69$ K and a water-feature strength of $S_{H_2O} = 0.079 \pm 0.072$. These results indicate that WASP-19b's dayside emission spectrum is consistent with those of other hot Jupiters with similar dayside temperatures (M. Mansfield et al. 2021; B. Edwards et al. 2023).

Our observations were unable to reliably constrain WASP-19b's atmospheric metallicity or C/O ratio, largely due to the limited wavelength range of our near-infrared data. Although there are multiple published studies of WASP-19b's optical transmission spectrum, these measurements are affected by WASP-19b's time-varying activity level. In the absence of ground-based stellar activity monitoring data, we were unable to mitigate the flux/transit depth offsets between the transmission spectrum data taken at different epochs. We conclude that this planet would benefit from additional transmission and emission spectroscopy measurements at longer wavelengths, which are less affected by stellar activity—for example, with JWST/NIRSpec PRISM. These observations would also improve our ability to measure the abundances of carbon-bearing molecules, necessary to obtain a robust constraint on the atmospheric C/O ratio.

## Acknowledgments

This research is based on observations made with the NASA/ESA Hubble Space Telescope obtained from the Space Telescope Science Institute (STScI), which is operated by the Association of Universities for Research in Astronomy, Inc., under NASA contract NAS 5-26555. These observations are associated with program GO-13431. The HST data used for the analysis in this paper can be found in MAST:10.17909/yjed-ka48. We acknowledge funding support from HST programs GO-13431 and GO-14767, and the Heising-Simons Foundation. We thank C. M. Huitson for leading the HST proposal on which this work is based; planning the observations; and providing feedback on the manuscript. M.W.M. and J.S. acknowledge support through the NASA Hubble Fellowship grant Nos. HST-HF2-51485.001-A and HST-HF2-51542.001-A, respectively, awarded by STScI.

*Software:* SciPy (P. Virtanen et al. 2020); NumPy (S. Van Der Walt et al. 2011); matplotlib (J. D. Hunter 2007); Astropy (Astropy Collaboration et al. 2013); batman (L. Kreidberg et al. 2015); emcee (D. Foreman-Mackey et al. 2013); PLATON (M. Zhang et al. 2019, 2020); dynesty (S. Koposov et al. 2023).

## ORCID iDs

Abigail A. Tumborang https://orcid.org/0009-0008-3005-0435
Jessica J. Spake https://orcid.org/0000-0002-5547-3775
Heather A. Knutson https://orcid.org/0000-0002-5375-4725
Megan Weiner Mansfield https://orcid.org/0000-0003-4241-7413
Kimberly Paragas https://orcid.org/0000-0003-0062-1168
Billy Edwards https://orcid.org/0000-0002-5494-3237
Tiffany Kataria https://orcid.org/0000-0003-3759-9080
Thomas M. Evans-Soma https://orcid.org/0000-0001-5442-1300
Nikole K. Lewis https://orcid.org/0000-0002-8507-1304
Gilda E. Ballester https://orcid.org/0000-0002-3891-7645

## References

Allard, F., Homeier, D., Freytag, B., & Sharp, C. M. 2012, in Atmospheres From Very Low-Mass Stars to Extrasolar Planets, ed. C. Reylé, C. Charbonnel, & M. Schultheis, Vol. 57 (Les Ulis: EDP Sciences), 3
Ambikasaran, S., Foreman-Mackey, D., Greengard, L., Hogg, D. W., & O'Neil, M. 2015, ITPAM, 38, 252
Anderson, D. R., Smith, A. M. S., Madhusudhan, N., et al. 2013, MNRAS, 430, 3422
Astropy Collaboration, Robitaille, T. P., Tollerud, E. J., et al. 2013, A&A, 558, A33
Barstow, J. K., Aigrain, S., Irwin, P. G. J., & Sing, D. K. 2017, ApJ, 834, 50
Batygin, K., & Stevenson, D. J. 2010, ApJL, 714, L238
Bean, J. L., Désert, J.-M., Seifahrt, A., et al. 2013, ApJ, 771, 108
Bell, T. J., & Cowan, N. B. 2018, ApJL, 857, L20
Berta, Z. K., Charbonneau, D., Désert, J.-M., et al. 2012, ApJ, 747, 35
Castelli, F., & Kurucz, R. L. 2003, in IAU Symp. 210, Modelling of Stellar Atmospheres, Poster Contributions, ed. N. Piskunov (San Francisco, CA: ASP), A20
Changeat, Q., Edwards, B., Al-Refaie, A. F., et al. 2022, ApJS, 260, 3
Charbonneau, D., Brown, T. M., Noyes, R. W., & Gilliland, R. L. 2002, ApJ, 568, 377
Claret, A. 2000, A&A, 363, 1081
Cortés-Zuleta, P., Rojo, P., Wang, S., et al. 2020, A&A, 636, A98
Coulombe, L.-P., Benneke, B., Challener, R., et al. 2023, Natur, 620, 292
Deming, D., Knutson, H., Agol, E., et al. 2011, ApJ, 726, 95
Deming, D., Seager, S., Richardson, L. J., & Harrington, J. 2005, Natur, 434, 740
Edwards, B., Changeat, Q., Tsiaras, A., et al. 2023, ApJS, 269, 46
Espinoza, N., Rackham, B. V., Jordán, A., et al. 2019, MNRAS, 482, 2065
Evans, T. M., Pont, F., Sing, D. K., et al. 2013, ApJL, 772, L16
Evans, T. M., Sing, D. K., Goyal, J. M., et al. 2018, AJ, 156, 283
Foreman-Mackey, D., Hogg, D. W., Lang, D., & Goodman, J. 2013, PASP, 125, 306
Fu, G., Deming, D., May, E., et al. 2021, AJ, 162, 271
Fu, G., Sing, D. K., Deming, D., et al. 2022, AJ, 163, 190
Gibson, N. P., Nikolov, N., Sing, D. K., et al. 2017, MNRAS, 467, 4591
Hebb, L., Collier-Cameron, A., Triaud, A., et al. 2009, ApJ, 708, 224
Hindle, A. W., Bushby, P. J., & Rogers, T. M. 2021a, ApJL, 916, L8
Hindle, A. W., Bushby, P. J., & Rogers, T. M. 2021b, ApJ, 922, 176
Hubeny, I., Burrows, A., & Sudarsky, D. 2003, ApJ, 594, 1011
Huitson, C. M., Sing, D. K., Pont, F., et al. 2013, MNRAS, 434, 3252
Hunter, J. D. 2007, CSE, 9, 90
Husser, T.-O., Wende von Berg, S., Dreizler, S., et al. 2013, A&A, 553, 9
Kataria, T., Sing, D. K., Lewis, N. K., et al. 2016, ApJ, 821, 9
Koposov, S., Speagle, J., Barbary, K., et al. 2023, joshspeagle/dynesty: v2.1.3, Zenodo, doi:10.5281/zenodo.8408702
Kreidberg, L., Bean, J. L., Désert, J.-M., et al. 2014a, ApJL, 793, L27
Kreidberg, L., Bean, J. L., Désert, J.-M., et al. 2014b, Natur, 505, 69
Kreidberg, L., Line, M. R., Bean, J. L., et al. 2015, arXiv:1504.05586
Lendl, M., Gillon, M., Queloz, D., et al. 2013, A&A, 552, A2
Line, M. R., Wolf, A. S., Zhang, X., et al. 2013, ApJ, 775, 137
Lothringer, J. D., Barman, T., & Koskinen, T. 2018, ApJ, 866, 27
Madhusudhan, N., Harrington, J., Stevenson, K. B., et al. 2011, Natur, 469, 64
Mandell, A. M., Haynes, K., Sinukoff, E., et al. 2013, ApJ, 779, 128
Mansfield, M., Line, M. R., Bean, J. L., et al. 2021, NatAs, 5, 1224
Menou, K. 2011, ApJL, 744, L16

Mikal-Evans, T., Sing, D. K., Goyal, J. M., et al. 2019, MNRAS, 488, 2222
Moses, J. I., Madhusudhan, N., Visscher, C., & Freedman, R. S. 2013, ApJ, 763, 25
Panwar, V., Désert, J.-M., Todorov, K. O., et al. 2022, MNRAS, 515, 5018
Parmentier, V., Line, M. R., Bean, J. L., et al. 2018, A&A, 617, A110
Parmentier, V., Showman, A. P., & Fortney, J. J. 2020, MNRAS, 501, 78
Parmentier, V., Showman, A. P., & Lian, Y. 2013, A&A, 558, A91
Perna, R., Menou, K., & Rauscher, E. 2010, ApJ, 724, 313
Pinhas, A., Madhusudhan, N., Gandhi, S., & MacDonald, R. 2019, MNRAS, 482, 1485
Rackham, B. V., Apai, D., & Giampapa, M. S. 2018, ApJ, 853, 122
Rajpurohit, A. S., Allard, F., Homeier, D., Mousis, O., & Rajpurohit, S. 2020, A&A, 642, A39
Rajpurohit, A. S., Reylé, C., Schultheis, M., et al. 2012, A&A, 545, A85
Redfield, S., Endl, M., Cochran, W. D., & Koesterke, L. 2008, in ASP Conf. Ser. 393, New Horizons in Astronomy, ed. A. Frebel et al. (San Francisco, CA: ASP), 259
Roman, M., & Rauscher, E. 2019, ApJ, 872, 1
Roman, M. T., Kempton, E. M.-R., Rauscher, E., et al. 2021, ApJ, 908, 101
Rustamkulov, Z., Sing, D. K., Mukherjee, S., et al. 2023, Natur, 614, 659
Sedaghati, E., Boffin, H. M. J., MacDonald, R. J., et al. 2017, Natur, 549, 238
Sedaghati, E., MacDonald, R. J., Casasayas-Barris, N., et al. 2021, MNRAS, 505, 435
Sing, D. K., Fortney, J. J., Nikolov, N., et al. 2016, Natur, 529, 59
Sing, D. K., Pont, F., Aigrain, S., et al. 2011, MNRAS, 416, 1443
Spake, J. J., Oklopčić, A., & Hillenbrand, L. A. 2021, AJ, 162, 284
Spiegel, D. S., Menou, K., & Scharf, C. A. 2009, ApJ, 691, 596
Tsiaras, A., Waldmann, I. P., Zingales, T., et al. 2018, AJ, 155, 156
Van Der Walt, S., Colbert, S. C., & Varoquaux, G. 2011, CSE, 13, 22
Virtanen, P., Gommers, R., Oliphant, T. E., et al. 2020, NatMe, 17, 261
Wakeford, H. R., Sing, D. K., Evans, T., Deming, D., & Mandell, A. 2016, ApJ, 819, 10
Wong, I., Benneke, B., Shporer, A., et al. 2020, AJ, 159, 104
Wong, I., Knutson, H. A., Kataria, T., et al. 2016, ApJ, 823, 122
Zhang, M., Chachan, Y., Kempton, E. M.-R., & Knutson, H. A. 2019, PASP, 131, 034501
Zhang, M., Chachan, Y., Kempton, E. M.-R., Knutson, H. A., & Chang, W. H. 2020, ApJ, 899, 27
Zhou, Y., Apai, D., Lew, B. W. P., & Schneider, G. 2017, AJ, 153, 243